\input epsf

\documentclass[10pt]{article}

\title{Pulse Shape Discrimination in the IGEX Experiment}

\begin{document}           % End of preamble and beginning of text.

\maketitle

\begin{center}
%List of authors
D. Gonz\'{a}lez$^{a}$, J. Morales$^{a}$\footnote{Corresponding
author: jmorales@posta.unizar.es}, S. Cebri\'{a}n$^{a}$, E.
Garc\'{\i}a$^{a}$, I.G. Irastorza$^{a}$\footnote{Present address:
CERN, EP Division, CH-1211 Geneva 23, Switzerland}, \\A.
Morales$^{a}$, A. Ortiz de Sol\'{o}rzano$^{a}$, J.
Puimed\'{o}n$^{a}$, M.L. Sarsa$^{a}$, J.A. Villar$^{a}$ \\ C.E.
Aalseth$^{b}$\footnote{Present address: Pacific Northwest National
Laboratory, Richland, WA 99352, USA}, F.T. Avignone III$^{b}$
\\ R.L. Brodzinski$^{c}$, W.K. Hensley$^{c}$, H.S. Miley$^{c}$, J.H. Reeves$^{c}$ \\
%\footnote{Corresponding author:
%amorales@posta.unizar.es},
I.V. Kirpichnikov$^{d}$, A.A. Klimenko$^{e}$ \\ S.B.
Osetrov$^{e}$,
%%S. Scopel$^{a}$,
A.A. Smolnikov$^{e}$, , A.A. Vasenko$^{e}$, S.I. Vasiliev$^{e}$\\
V.S. Pogosov$^{f}$, A.G. Tamanyan$^{f}$
\end{center}

\begin{center}
\begin{em}
$^{a}$Laboratory of Nuclear and High Energy Physics, University of
Zaragoza, \\ 50009 Zaragoza, Spain
\\
$^{b}$University of South Carolina, Columbia, South Carolina 29208
USA
\\
$^{c}$Pacific Northwest National Laboratory, Richland, Washington
99352 USA
\\
$^{d}$Institute for Theoretical and Experimental Physics, 117 259
Moscow, Russia
\\
$^{e}$Institute for Nuclear Research, Baksan Neutrino
Observatory,\\  361 609 Neutrino, Russia
\\
$^{f}$Yerevan Physical Institute, 375 036 Yerevan, Armenia \\
\end{em}
\end{center}

\begin{abstract}
The IGEX experiment has been operating enriched germanium
detectors in the Canfranc Underground Laboratory (Spain) in a
search for the neutrinoless double decay of $^{76}$Ge. The
implementation of Pulse Shape Discrimination techniques to reduce
the radioactive background is described in detail. This analysis
has been applied to a fraction of the IGEX data, leading to a
rejection of $\sim$60 \% of their background, in the region of
interest (from 2 to 2.5 MeV), down to $\sim 0.09$ c/(keV kg y).
\end{abstract}

\section{Introduction}

The nuclear Double Beta Decay (DBD) is an unique laboratory to
investigate the nature and properties of the neutrino
\cite{reviews,morales}. The neutrinoless decay mode, if it exists,
would provide an unambiguous evidence of the Majorana nature of
the neutrino, its non-zero mass, and the non-conservation of
lepton number. After the definitive confirmation that neutrinos
have indeed non-zero mass, as the solar and atmospheric neutrino
oscillation results imply \cite{toshito,ahmad}, the neutrinoless
Double Beta Decay (DBD) has become a most relevant subject of
research because it is a process able to provide, in a relatively
short time, the neutrino mass scale and its hierarchy pattern. The
current best sensitivity limits of the effective Majorana electron
neutrino mass -derived from the neutrinoless half-life lower
bound- stands around $\langle m_{\nu} \rangle \leq 300-1300$ meV
\cite{igex,klapdor} (the dispersion range is due to the
uncertainties in the evaluation of the nuclear matrix elements),
whereas most of the neutrino mass models indicate that the mass
region where the Majorana nature of the neutrino would be resolved
lies two orders of magnitude below ($\sim$ a few meV).
%Which would require substatial improvements in
%the relevant experimental parameters? Number of 2$\beta$ emitter
%nuclei,
To achieve the sensitivity required for such new objectives, it
will require a large number of double beta emitter nuclei, a very
low background and a sharp energy resolution in the Q-value
region, and/or methods to disentangle signal from noise.

A typical example of this type of search is IGEX. The
International Germanium EXperiment (IGEX) has been running in the
Canfranc Underground Laboratory at a depth of 2450 m.w.e. in a
search for the neutrinoless Double Beta Decay.
%A long-term stability has
%been proven for gain and energy resolution.
Details and results of the experiment can be found in ref.
\cite{igex}.
%uses unique cryostat
%technology, ultralow background material, archaeological lead
%shielding and a pulse shape analysis method described in the
%following. IGEX
Three Germanium detectors (RG1, RG2 and RG3), of $\sim$2 kg each,
enriched to 86\% in $^{76}$Ge were used.

%The Pulse Shape Discrimination (PSD) is a technique to reject
%radioactive background. The number of points in which an
%interaction occurs (and where the energy is released) is reflected
%to some extent in the shape of the electric pulse delivered by the
%detector; in particular, in the number of lobes the pulse has.
%Events in which all the energy is deposited in only one point (the
%so-called Single-Site Events, SSE) may be due to a photoelectric
%absortion, a single Compton interaction, a pair creation with
%double escape as well as to Double Beta Decays, since the emitted
%electrons slow down almost immediately in the surrounding medium.
%On the other hand, Multi-Site Events (MSE) could be produced by
%different processes: multi-Compton scattering, Compton $+$
%photoelectric interaction, or other typical processes originated
%by the radioactive background. As a first conclusion, a
%neutrinoless DBD event should be SSE, while radioactive background
%gives mainly MSE.

This paper presents the development and application of one
technique used to reduce part of the radioactive background by
discriminating it from the expected signal by comparison of the
shape of the pulses (PSD) of both types of events. As an example,
the method has been applied to the data recorded by two Ge
detectors of the IGEX Collaboration, which has produced one of the
two best current sensitivity limits for the Majorana neutrino mass
parameter \cite{igex,klapdor}.

The rationale for PSD is quite simple: in large intrinsic Ge
detectors, the charge carriers take 300~-~500 ns to reach their
respective electrodes. These drift times are long enough for the
current pulses to be recorded at a sufficient sampling rate. The
current pulse contributions from electrons and holes are
displacement currents, and therefore dependent on their
instantaneous velocities and locations. Accordingly, events
occurring at a single site ($\beta\beta$-decay events for example)
have associated current pulse characteristics which reflect the
position in the crystal where the event occurred. More
importantly, these single-site events (SSE) frequently have pulse
shapes that differ significantly from those due to the background
events that produce electron-hole pairs at several sites by
multi-Compton-scattering process, for example (the so-called
Multi-Site Events (MSE)). Consequently, pulse-shape analysis can
be used to distinguish between these two types of energy
depositions:
%DBD events will deposit energy at a single site in a
%detector while background events will deposit energy at several
%sites (in particular in energy regions where Compton processes are
%important).
DBD events belong to the SSE class of events and will deposit
energy at a single site in the detector while most of the
background events belong to the MSE class of events and will
deposit energy at several sites.

The IGEX detectors have modified preamplifier electronics to route
and record the current pulses at a very early stage of
preamplification, thus producing unique high-bandwidth pulse shape
signals. Furthermore, to develop PSD techniques it would be highly
desirable to obtain an earlier signal, even before it passes
through the few unavoidable electronic components at the first
stage of the detector preamplifier, resembling as much as possible
the displacement current of the detector. This allows the
development of algorithms that do not depend strongly on the
preamplifier electronics in use. To this end, the transfer
function of the preamplifier and associated front-end stage has
been measured for each detector. This allows the reconstruction of
the displacement current and an easy comparison to computed pulse
shapes.

 %Our models of the structure of the current pulse
%reveal that single-site events will exhibit only one or two
%features, or ``lobes'', in more than 97\% of the cases. Multiple
%site events will most often exhibit more than two lobes.

%One PSD technique is to reject pulses having more than two
%significant lobes or peaks. To detect lobes, a ``Mexican-hat''
%filter of the proper width is applied to the pulse. This robust
%method is nearly model-independent.  Some multiple-site events may
%show only one or two lobes and will not be rejected by this
%technique.  Use of this PSD method results in the rejection of
%60\%--80\% of the IGEX background in the energy interval
%2.0--2.5~MeV, down to less than $\sim0.07$c/keV.kg.y.

%powerful tool

The paper is organized as follows: in Section \ref{ecp}, the
properties of the digitized experimental pulses are shown and the
work performed to understand and reproduce their shapes is
presented. In Section \ref{methods}, the PSD method is described.
Finally, Section \ref{results} displays the results of this
analysis for the IGEX data.

%The implementation of PSD techniques in the IGEX experiment is
%reviewed in this paper. In section \ref{ecp}, the properties of
%the digitized experimental pulses are shown and the work performed
%to understand and reproduce their shapes is presented. In section
%\ref{methods}, different methods to apply PSD are described.
%Finally, section \ref{results} contains the results of this
%analysis for the IGEX data.

\section{Experimental and calculated pulses}
\label{ecp}

\begin{figure}[tb]
\centerline{ \fboxrule=0cm
 \fboxsep=1cm
  \fbox{
\epsfxsize=12cm%estaba 15
  \epsfbox{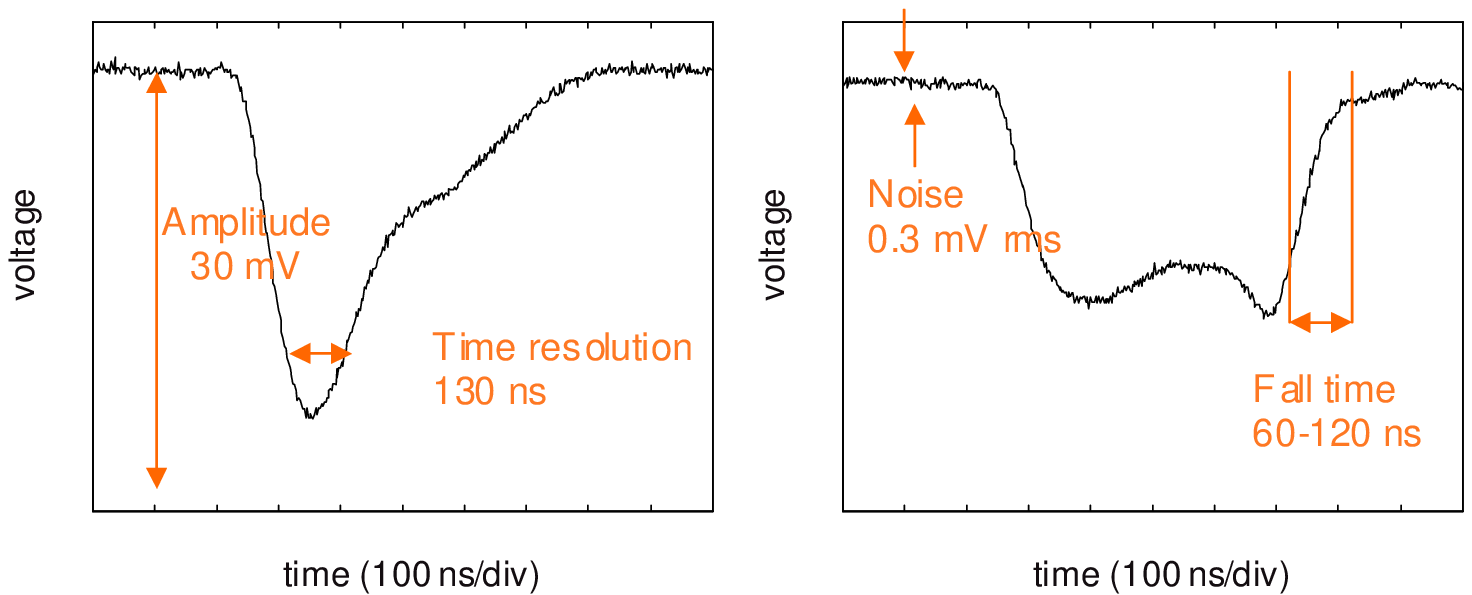}}}
\begin{center}
{\caption {Main features of the digitized experimental pulses.}
\label{properties}}
\end{center}
\end{figure}

\begin{figure}[tb]
\centerline{ \fboxrule=0cm
 \fboxsep=0cm
  \fbox{
\epsfxsize=15cm
  \epsfbox{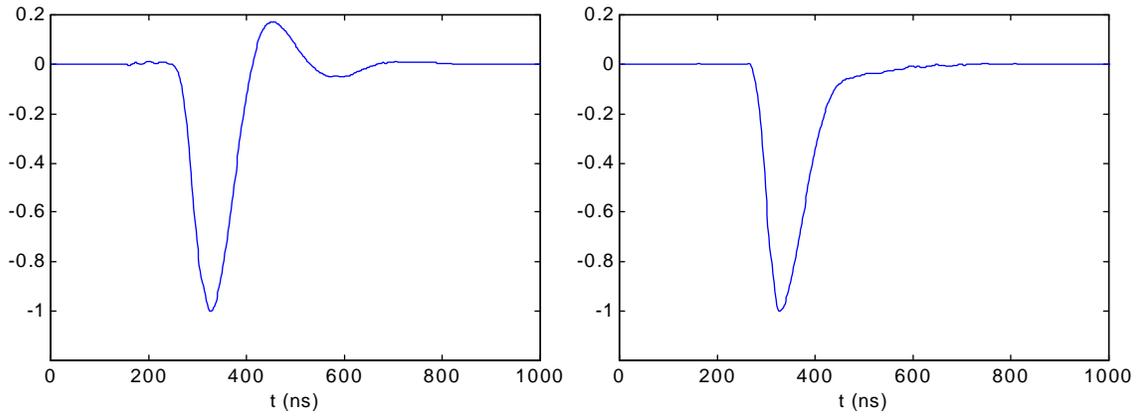}}}
\begin{center}
{\caption {Transfer functions of the preamplifiers for detectors
RG2 (on the left) and RG3 (on the right).} \label{transfer}}
\end{center}
\end{figure}

    The data acquisition system of the IGEX experiment is based on
standard NIM electronics, each Ge detector having an independent
electronic chain. Preamplifiers were modified for pulse shape
analysis and each preamplifier fast-pulse output is routed to a
LeCroy 9362 digital oscilloscope (800 MHz analog bandwith). The
digitized pulse signal covers a total time of 1 $\mu$s using 500
points; it is worth noting that the time resolution, of about 100
ns (as inferred from the width of the peaked features or the fall
time), limits the ability to resolve nearby features in the pulse
such as lobes or discontinuities characteristic of a multiple-site
interaction signal. Figure \ref{properties} shows the main
features of the digitized pulses. These output pulses are taken at
the very first stage of the amplification chain, but even so,
there is an unavoidable instrumental distortion due to the
preamplifier. This has been studied to determine its transfer
function $h(t)$. To take into account this distortion in the pulse
shape analysis, either the calculated pulse $i(t)$ is folded with
the transfer function,
\begin{equation}
o(t)=\int_{-\infty}^{+\infty}i(\tau)h(\tau-t)d\tau
\end{equation}
or the experimental signal unfolded. The latter allows the
recovery of some information lost because of the instrumental
distortion \cite{br88}.
%m\'{a}s sobre deconvoluci\'{o}n, en espacio de frecuencias, corte...
The transfer function of the preamplifiers, $h(t)$, depicted in
figure \ref{transfer}, has been directly measured as the response
of the preamplifier for a narrow $\delta$-like signal. Studies
were also made following several methods including analog
simulation of the preamplifier circuits and the analysis of the
shapes of selected populations of experimental pulses. It was
observed that the shape of the rise and the fall of the folded
pulses is the same that the shape of the transfer function for
those events in which the energy is released in some particular
regions of the crystal; in particular, the left side of the
transfer function can be deduced by studying pulses of events in
the inner and lower part of the crystal, while the right side is
derived from those produced in the outer and upper region.
Measurements and estimates for the transfer function are found to
be in quite good agreement.

    The pulse shapes of the output signals can be reproduced numerically.
An energy deposition in a Ge crystal produces a proportional
number of electron-hole pairs, which move towards the electrodes.
The induced current $i$, and consequently the electric pulse taken
from the detector, is the sum of the contributions due to each
type of charge carriers:
\begin{equation}
i(t)=i_{\mathrm{e}}(t)+i_{\mathrm{h}}(t)
\end{equation}
The current is calculated as \cite{ra88}:
\begin{equation}
i_{\mathrm{e(h)}}=-q_{\mathrm{e(h)}}\overrightarrow{E}_{\mathrm{w}}\overrightarrow{v_{\mathrm{e(h)}}}(\overrightarrow{E})
\end{equation}
%explicar el origen de esta expresion?
\noindent where $\overrightarrow{E}_{\mathrm{w}}$ is the so-called
weighting field\footnote{The weighting field is the field that
would appear in the crystal if a unity voltage was applied to the
sensor electrode (assuming no impurity in the crystal)}. In
principle, the drift velocity $v$ of charge carriers is
proportional to the electric field $E$: $v(r)=\mu E(r)$, where
$\mu$ is the mobility in the material medium. However, at high
values of the electric field, a saturation velocity is reached.
The dependence of the velocity on the electric field has been
obtained empirically \cite{cm75}:
%gr\'{a}fico de velocidades?
\begin{equation}
v(E)=\frac{\mu
E}{(1+(\frac{E}{E_{\mathrm{sat}}})^{\beta})^{1/\beta}} \label{ve}
\end{equation}

The values commonly used for the parameters involved in expression
\ref{ve} (the mobility $\mu$, the numerical parameter $\beta$ and
the electric flield for saturation $E_{\mathrm{sat}}$) are
summarized in table \ref{parve} \cite{rb82}.

\begin{table}[htb]
\centering \caption {Values commonly considered for the parameters
involved in the empirical dependence of the velocity of the charge
carriers on the electric field.} \label{parve} \vskip 0.5 cm
\begin{tabular}{|c|c|c|} \hline

 &   electrons &  holes \\ \hline
$E_{\mathrm{sat}}$(V/cm) & 275 & 210.5 \\ \hline

$\mu$ (cm$^{2}$/Vs) & 36000 &  42000 \\ \hline

$\beta$ & 1.32 & 1.36 \\ \hline

\end{tabular}
\end{table}

The solution for the electric field
($\overrightarrow{E}(\overrightarrow{r})=\overrightarrow{\nabla}\phi(\overrightarrow{r})$)
in the crystal is derived from the Poisson equation, applied for
the depleted and non-depleted regions:

\begin{equation}
\nabla^{2}\phi(\overrightarrow{r})=-\frac{\rho(\overrightarrow{r})}{\epsilon},
\, \mathrm{depleted}
\end{equation}
\begin{equation}
\nabla^{2}\phi(\overrightarrow{r})=0, \, \mathrm{non-depleted}
\end{equation}
\begin{equation}
\phi(\overrightarrow{r}_{{\mathrm
int}})-\phi(\overrightarrow{r}_{{\mathrm ext}})=V_{0}
\label{bound}
\end{equation}

\begin{figure}[tb]
\centerline{ \fboxrule=0cm
 \fboxsep=1cm
  \fbox{
\epsfxsize=12cm %estaban 0, 15 cm
  \epsfbox{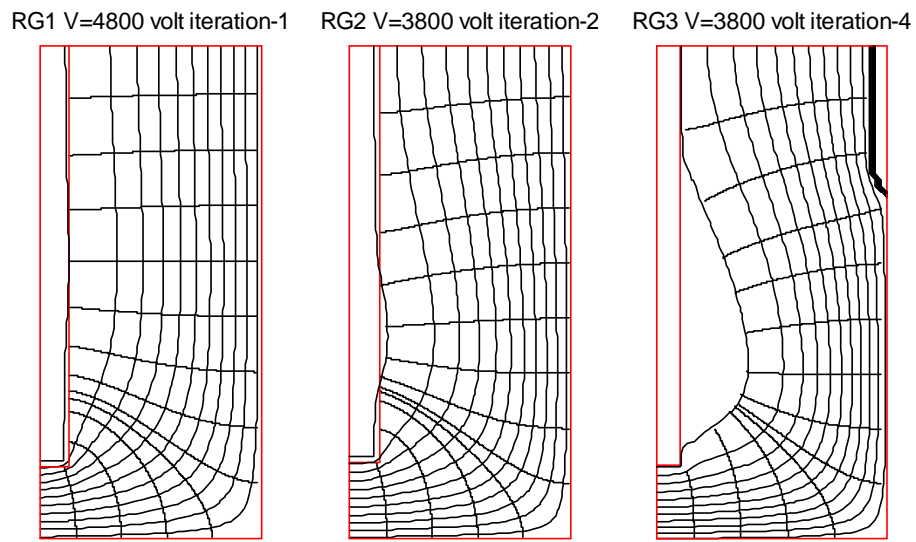}}}
\begin{center}
{\caption {Electric field for detectors RG1, RG2 and RG3. The
operation voltage and the number of iterations necessary to derive
the field (see text) are shown for each detector.} \label{efield}}
\end{center}
\end{figure}

    The electric field depends on the geometry of the crystal (true coaxial or closed-end), the
supplied voltage $V_{0}$ and on the residual space charge density
$\rho$. The parameters $r_{\mathrm{int}}$ and $r_{\mathrm{ext}}$
correspond to the locations of the internal and external
electrodes of the detector. A solution (depicted in figure
\ref{efield}) has been obtained for the cases of depleted and
non-fully depleted, closed-end crystals of the IGEX experiment by
using an iterative method of calculation. For a non-fully depleted
detector, the boundary between the depleted and non-depleted
regions in the crystal is unknown a priori. An invalid solution is
obtained when using the boundary condition of eq. (\ref{bound}),
because a region with a field having inverse direction appears
(which means that this region is not depleted). Therefore, the
boundary condition is changed by using the deduced limit for the
depleted region instead of $r_{\mathrm{int}}$; then, a new
solution is found for the electric field and, consequently, a new
limit for the non-depleted region is obtained. This procedure is
repeated until the solution does not change significantly between
two successive iterations.

\begin{figure}[tb]
\centerline{ \fboxrule=0cm
 \fboxsep=0cm
  \fbox{
\epsfxsize=10cm %estaba 14, luego 12
  \epsfbox{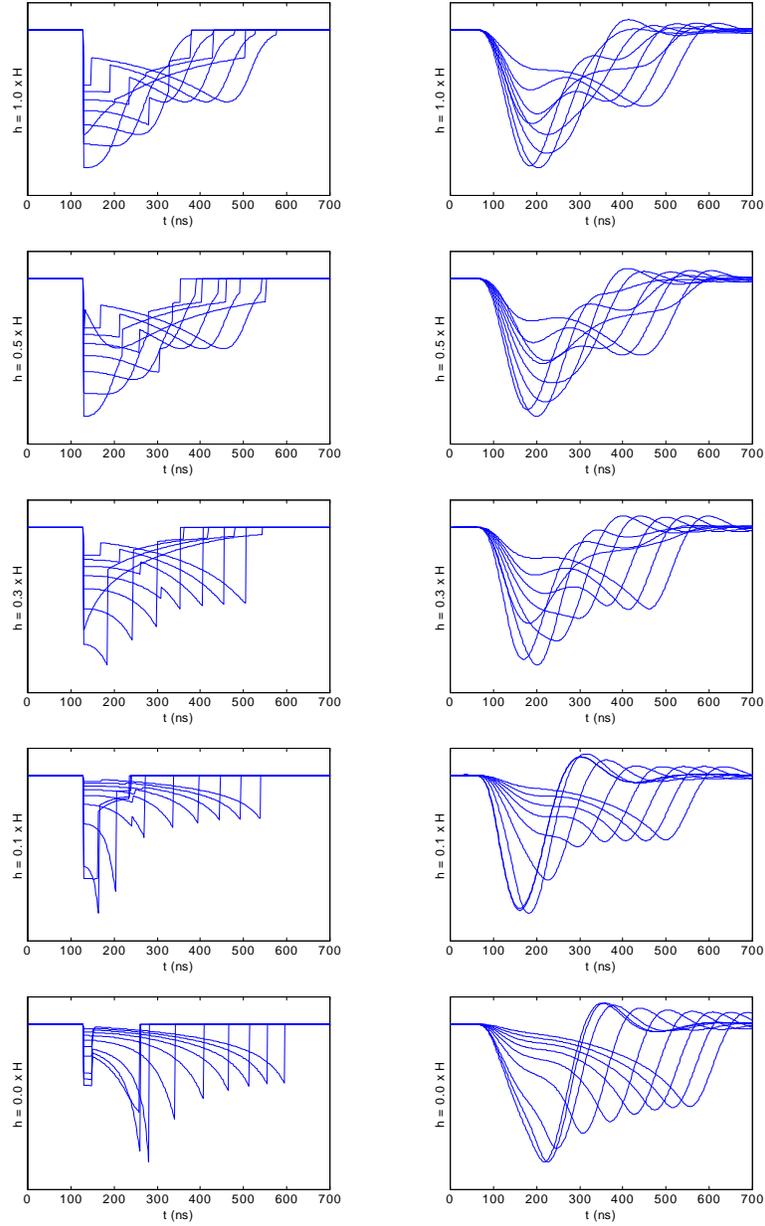}}}
\begin{center}
{\caption {Calculated pulse shapes for SSE in detector RG2.
Different radial positions have been considered in each plot for
several vertical coordinates h (H is the height of the crystal).
On the right, the instrumental distortion has been taken into
account.} \label{pulses}}
\end{center}
\end{figure}

    Once the electric field is known, the pulse shapes can be
calculated. Some examples for SSE are shown in figure
\ref{pulses}. Different radial positions have been considered in
each plot for several vertical coordinates; the effect of the
instrumental distortion is shown on the right plots. To reproduce
MSE pulses, the shapes due to each individual interaction should
be properly added weighted by the fractional amount of energy
released in the crystal, obtained by Monte Carlo simulation.

\section{PSD Method}
\label{methods}

%    Three different PSD methods to reject background have been taken into consideration.
%The first one is based on the time of rise and fall of the pulses.
%Figure \ref{risefall} presents the distribution of these times for
%calculated SSE and for background events, in RG2 and RG3
%detectors. By comparing these distributions, the criterion to
%reject a pulse from the SSE set can be fixed: an event will be
%eliminated if both the fall and rise times are greater than 100
%ns.

\begin{figure}[tb]
\centerline{ \fboxrule=0cm
 \fboxsep=0cm
  \fbox{
\epsfxsize=15cm
  \epsfbox{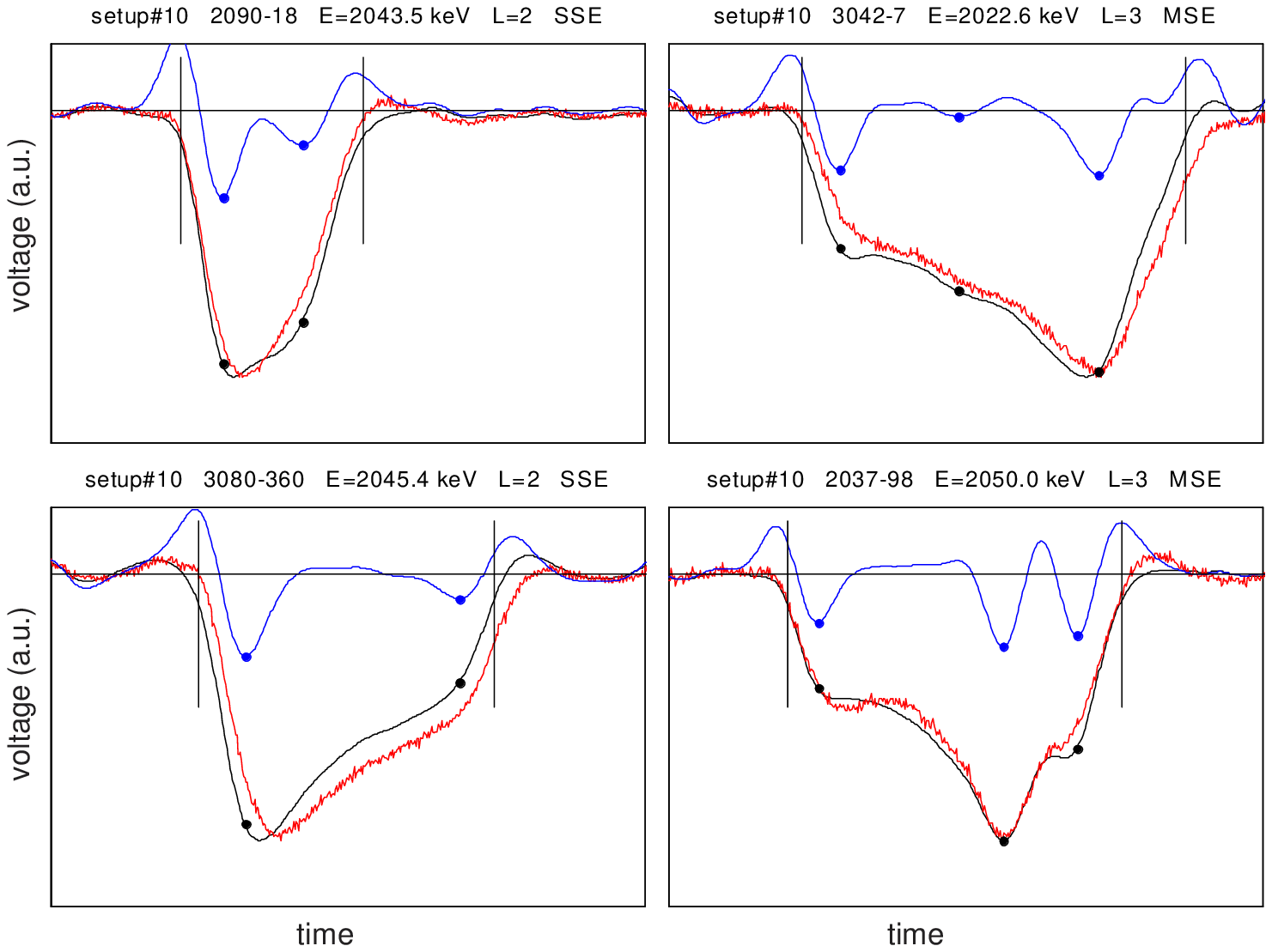}}}
\begin{center}
{\caption {Examples of the effect of applying the "mexican-hat"
filter to detect significant lobes in the digitized pulses.
%Thick lines correspond to unfolded digitized pulses and thin lines show
%filtered pulses.
Events on the left are accepted (having two lobes) while those on
the right are rejected (having three lobes).} \label{lobes}}
\end{center}
\end{figure}

    The PSD method we have used consists in counting the number of lobes of the
pulses and rejecting those events having more than two significant
lobes or peaks. A SSE pulse is expected to have at most two lobes,
one due to electrons and the other due to holes. Experimental
pulses are first unfolded using the transfer function of the
preamplifier. Then, to detect lobes a "mexican-hat" filter $F$ of
the proper width $\sigma$ is applied to the pulse. In fact, this
filter is the second derivative of a gaussian:
\begin{equation}
F(t)=\frac{\sigma^{2}-t^{2}}{\sigma^{4}}\times\exp(-\frac{t^{2}}{2\sigma^{2}})
\end{equation}
and the filtered signal has a null mean value where there is no
lobe in the original signal and a peak where a lobe is present.
%anchura ???? (FWHM=60 ns) discusi\'{o}n en notas de David .
Therefore, it is straightforward to reject all the events having
more than two lobes.
%This robust method is nearly
%model-independent, it does not rely directly on PS calculations.
%It is conservative, since some mse may show only 1 or 2 lobes and
%therefore will not be rejected.
Figure \ref{lobes} shows the results of applying this method to
four different pulses.

%This robust method is nearly model-independent. Its effectiveness
%has been evaluated on calculated SSE. A test population was
%generated and the method was applied to pulses folded or unfolded,
%following the same procedure used with the experimental pulses.
%The plot in figure \ref{howitworks} shows the fraction of
%misidentified SSE (i.e., events that appear to have 3 or more
%lobes) for the 3 different detectors, as a function of the
%threshold for analyzing the curvature of filtered pulses at lobe
%positions. This curvature is (as well as the width of the filter)
%a free parameter of the method; a threshold value has to be fixed
%to accept or not possible lobes. This plot has allowed to fix a
%safe value for this curvature, such that the error in the
%filtering is less than 3\%. This means that, according to our
%models of the structure of the current pulse, SSE will exhibit
%only one or two features, or ``lobes'', in more than 97\% of the
%cases.
%%explicar m\'{a}s los problemas de los unfolded, frecuencia..
%On the other hand, MSE will most often exhibit more than two
%lobes. Those cases which may show only one or two lobes (because
%the separation -in the radial coordinate- between the interaction
%points is not enough to make them distinguishable or because of
%extremely unbalanced fractions of energy deposited in each point)
%will not be rejected by this technique.

\begin{figure}[tb]
\centerline{ \fboxrule=0cm
 \fboxsep=0cm
  \fbox{
\epsfxsize=15cm
  \epsfbox{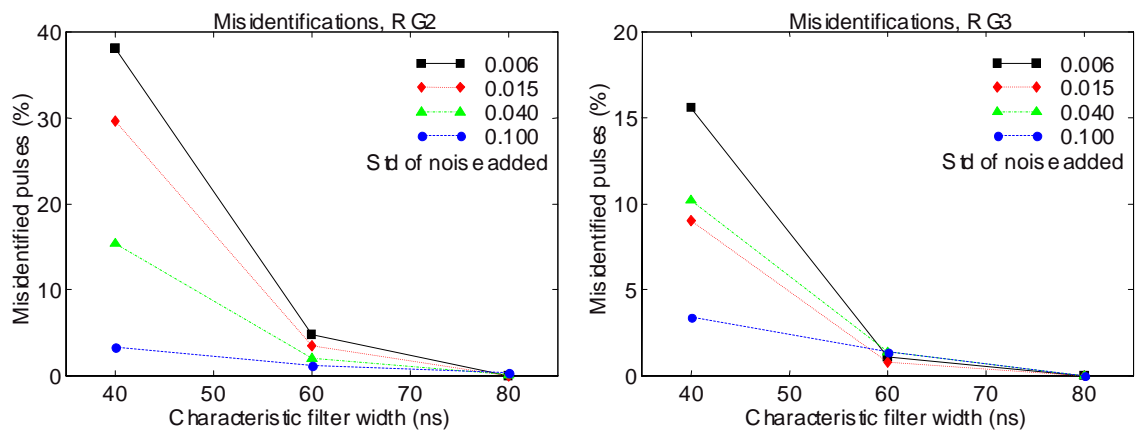}}}
\begin{center}
{\caption {%Fraction of misidentified calculated SSE as a function
%of the threshold for analyzing the filter curvature at lobe
%positions (see text).
Fraction of misidentified calculated SSE pulses as a function of
the characteristic widht of the analyzing filter and the amount of
noise added to the pulse (see text).} \label{howitworks}}
\end{center}
\end{figure}

This robust method is nearly model-independent. Its effectiveness
has been evaluated on calculated SSE. A test population of 2000
SSE pulses was generated for each detector. The locations were
randomly chosen, uniformly distributed in the volume of the
crystals. Each calculated pulse was folded with the proper
preamplifier transfer function, then scaled to unit height, and a
variable amount of gaussian noise was added, to reproduce
experimental pulses of different energies and noise levels.
Finally, the number of lobes of each pulse was obtained by
applying "mexican-hat" filters derived from gaussians of different
widths. After the analysis, the fraction of pulses with three o
more lobes was retained. This is an estimation of the expected
number of misidentified SSE pulses with this technique. The
results are shown in figure \ref{howitworks}. When applied to a
mixed set of SSE and MSE pulses, to obtain the best discrimination
results, a narrow filter should be preferred over a broad filter,
but it would produce a large fraction of misidentifications.
Conversely, a safer, broader filter would not find as many lobes
as a narrower filter, and its discrimination power would be
smaller. As a compromise, we choose the filter with a
characteristic width of 60 ns, thus keeping the misidentification
error for the calculated SSE pulses under 5\% for all pulses in
the whole noise range considered. Notice that those MSE in which
the separation of the interactions in the (r,z) plane is too small
or the amounts of energy deposited at each point are extremely
unbalanced will not be rejected by this technique.

%The PSD could be a powerful tool to reject background, but the
%efficiency of a clear identification of MSE is limited by several
%factors:

%    The third PSD method is based on the comparison of the
%digitized experimental pulses with a collection of well-known SSE
%pulses, either obtained experimentally \cite{hellmig} or
%numerically. The difference between the experimental pulse and the
%closest template can be so quantified, either by comparing areas
%or amplitudes; then, the events differing from the template more
%than a fixed cut value are rejected.
%Figure \ref{comparison} shows the comparison between experimental
%pulses and calculated SSE already folded, i.e., distorted. As
%already said, the comparison may be done also between unfolded
%experimental pulses and calculated not-degraded SSE. In principle,
%more information is available in this second way, but a frequency
%cut must be fixed in the process of unfolding the experimental
%pulse. Although a good knowledge of the pulse shapes has been
%achieved, this method has not been finally used in the following
%chosen application to the IGEX recorded spectra, because of its
%dependence on the model of pulse reconstruction.

\begin{figure}[tb]
\centerline{ \fboxrule=0cm
 \fboxsep=0cm
  \fbox{
\epsfxsize=12cm %estaba 15, luego 13
  \epsfbox{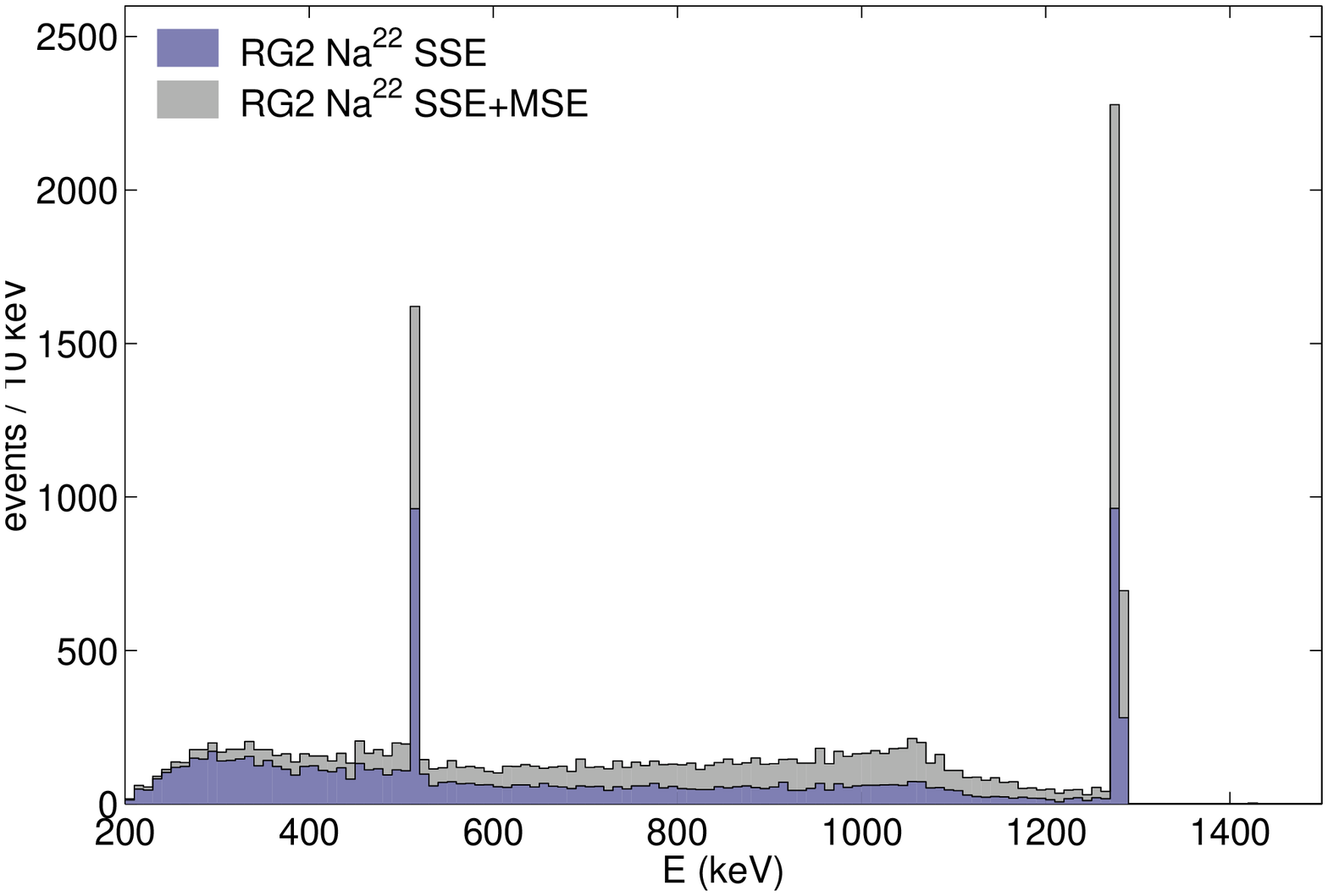}}}
\centerline{ \fboxrule=0cm
 \fboxsep=0cm
  \fbox{
\epsfxsize=12cm %estaba 15, luego 13
  \epsfbox{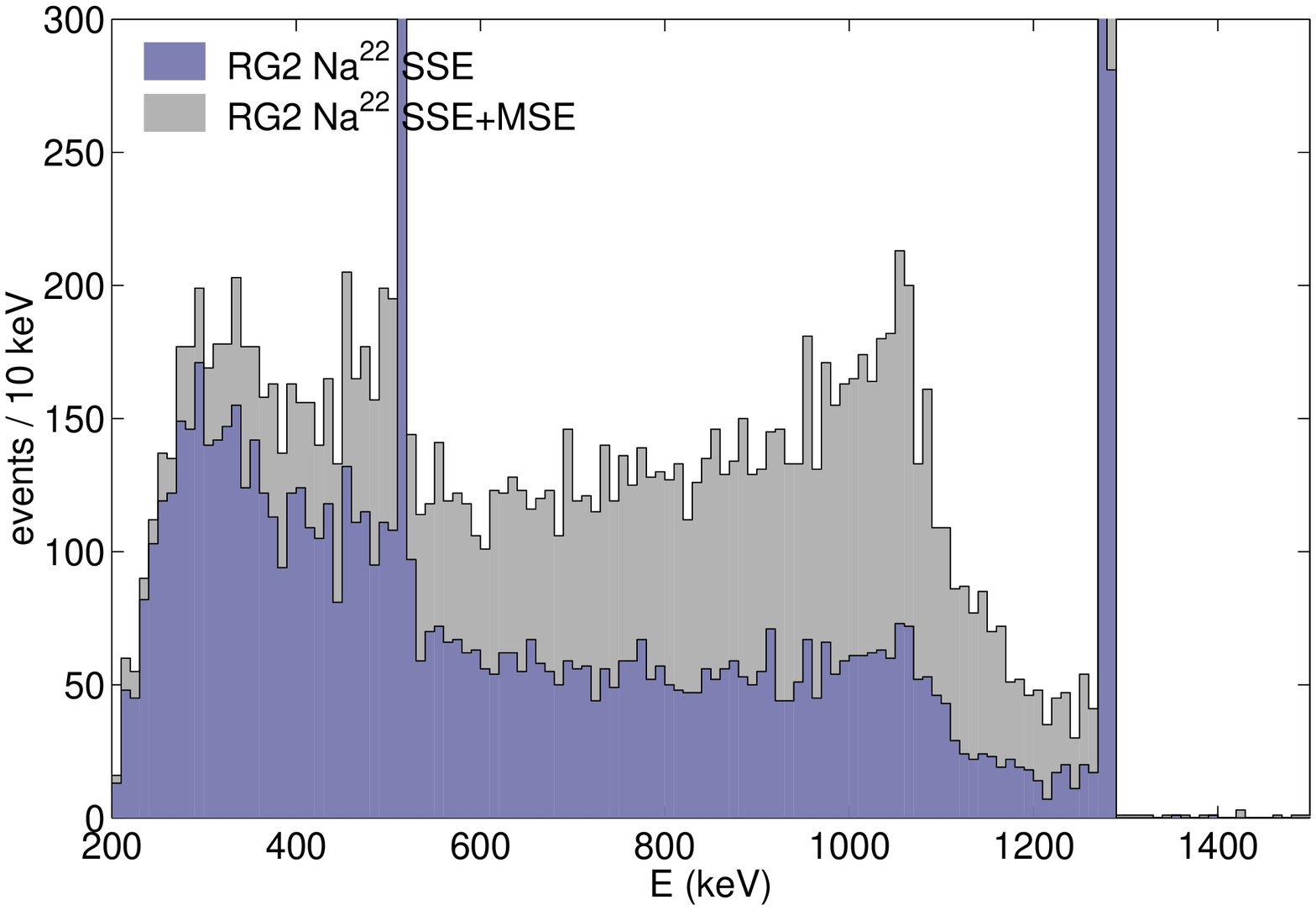}}}
\begin{center}
{\caption {Energy spectra before and after the PSD based on the
counting of the number of lobes for a $^{22}$Na calibration and
for detector RG2. A zoom of the plot at the top is shown at the
bottom.} \label{na}}
\end{center}
\end{figure}

\begin{figure}[tb]
\centerline{ \fboxrule=0cm
 \fboxsep=0cm
  \fbox{
\epsfxsize=12cm %estaba 15, luego 13
  \epsfbox{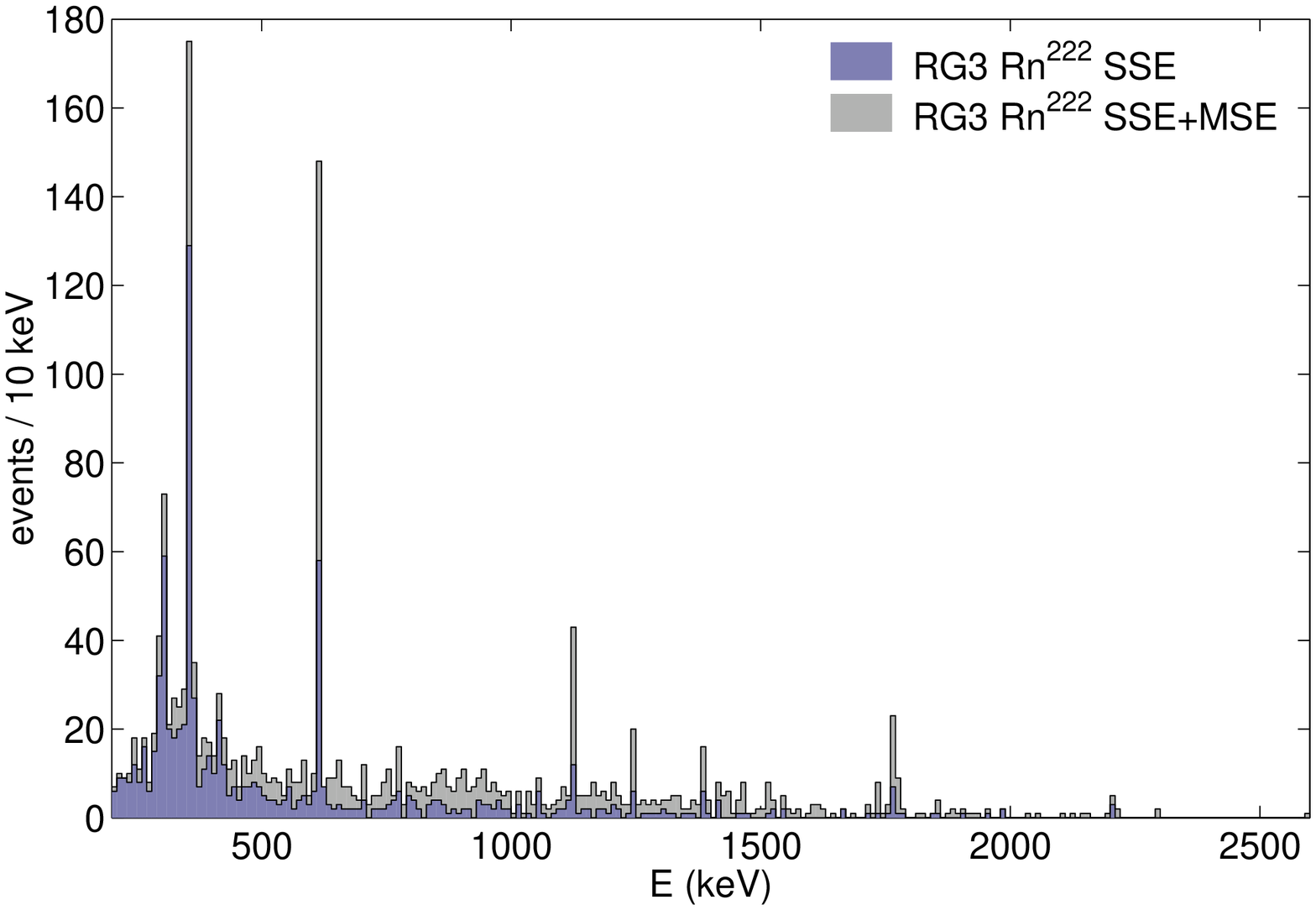}}}
  \centerline{ \fboxrule=0cm
 \fboxsep=0cm
  \fbox{
\epsfxsize=12cm %estaba 15, luego 13
  \epsfbox{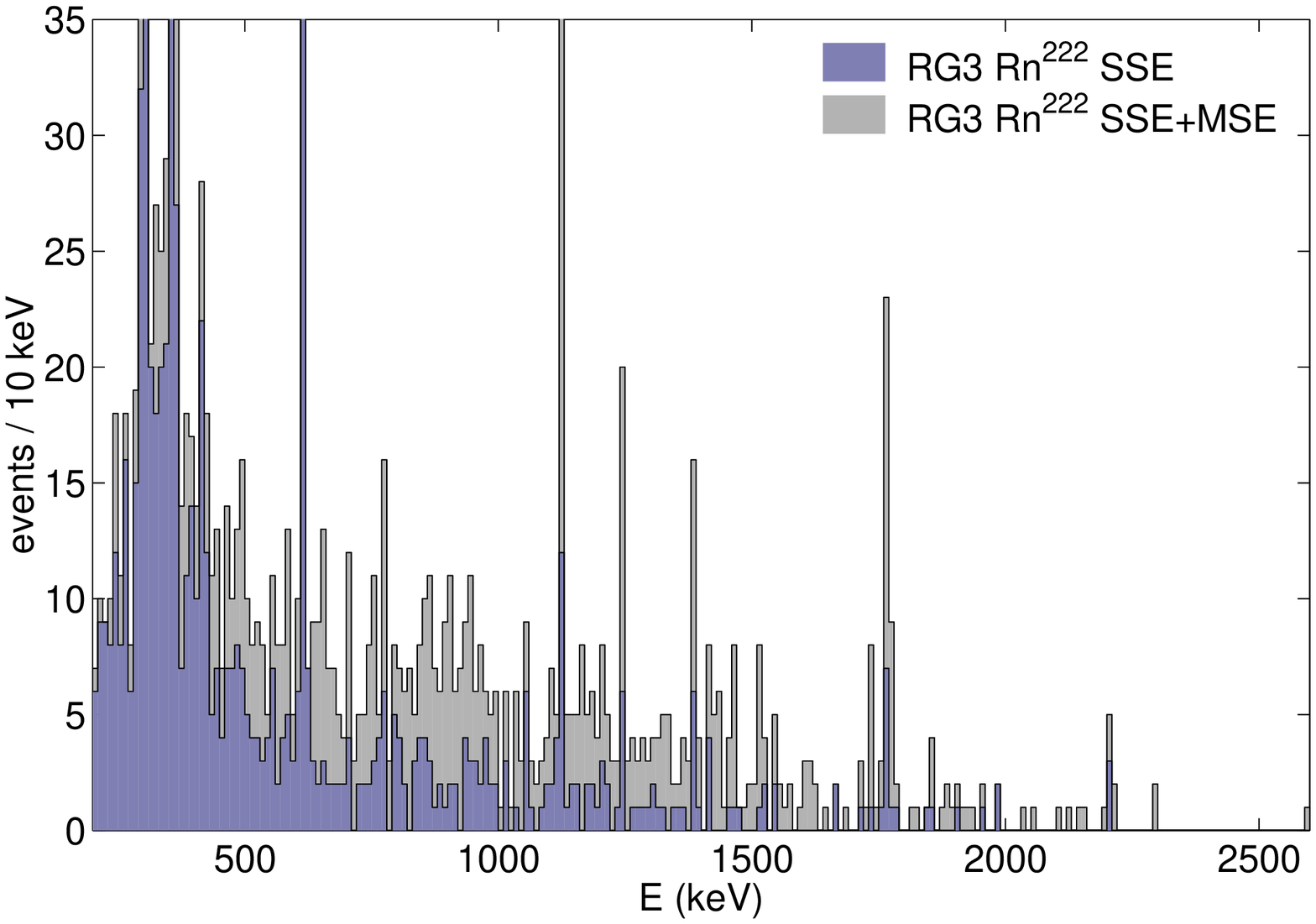}}}
\begin{center}
{\caption {Energy spectra before and after the PSD based on the
counting of the number of lobes for radon data and for detector
RG3. A zoom of the plot at the top is shown at the bottom.}
\label{radon}}
\end{center}
\end{figure}

\begin{figure}[tb]
\centerline{ \fboxrule=0cm
 \fboxsep=0cm
  \fbox{
\epsfxsize=12cm %estaba 15, luego 13
  \epsfbox{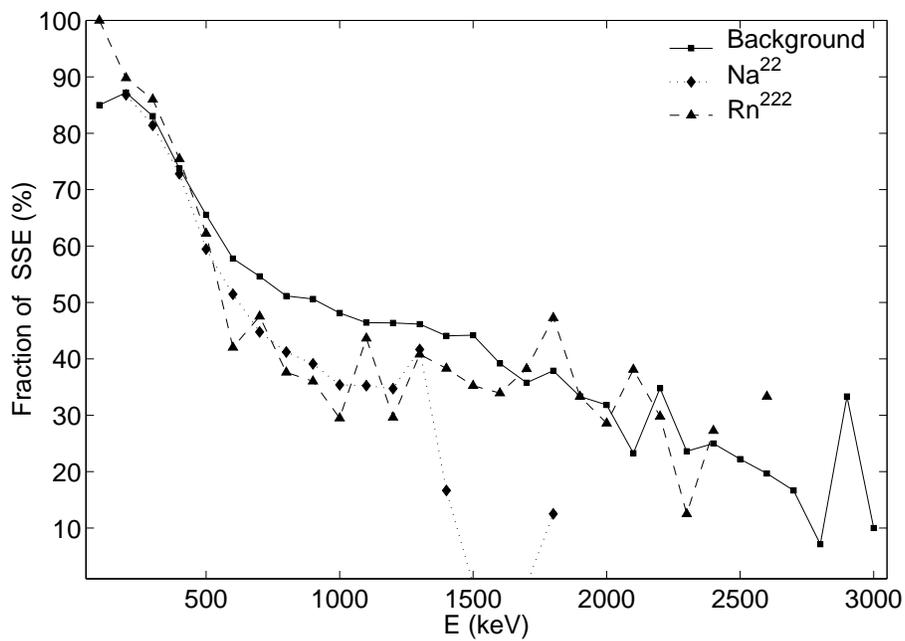}}}
  \centerline{ \fboxrule=0cm
 \fboxsep=0cm
  \fbox{
\epsfxsize=12cm %estaba 15, luego 13
  \epsfbox{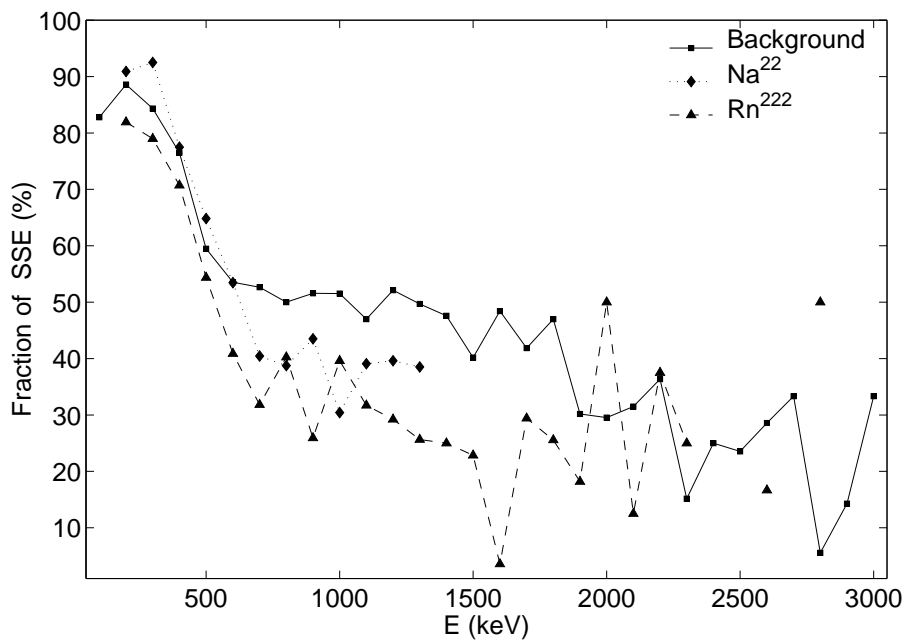}}}
\begin{center}
{\caption {Comparison of the percentage of SSE events identified
for a $^{22}$Na calibration, radon data and background, for both
detectors RG2 (top) and RG3 (bottom).} \label{percentage}}
\end{center}
\end{figure}

This method of counting the number of lobes of the pulses has been
applied also to a $^{22}$Na calibration spectrum and to a set of
data taken following a large intrusion of radon in the shielding.
Fig. \ref{na} illustrates the reduction in the case of $^{22}$Na
spectrum for detector RG2 and Fig. \ref{radon} the case of radon
for detector RG3. A comparison of the results of the method for
the cases of background, $^{22}$Na calibration and radon is shown
in Fig. \ref{percentage}.

\section{Results}
\label{results}

\begin{figure}[tb]
\centerline{ \fboxrule=0cm
 \fboxsep=0cm
  \fbox{
\epsfxsize=12cm %estaba 15, luego 13
  \epsfbox{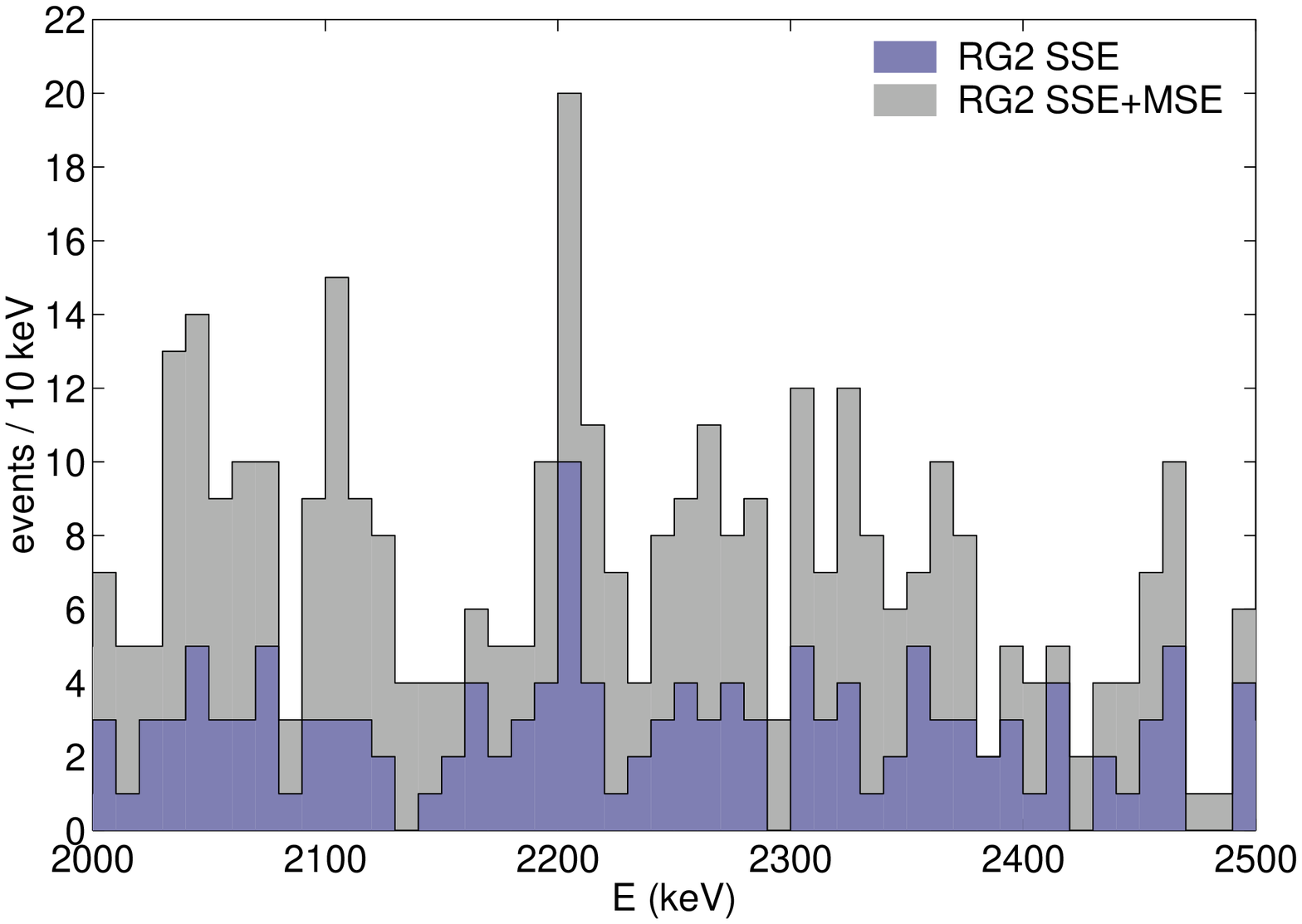}}}
\centerline{ \fboxrule=0cm
 \fboxsep=0cm
  \fbox{
\epsfxsize=12cm %estaba 15, luego 13
  \epsfbox{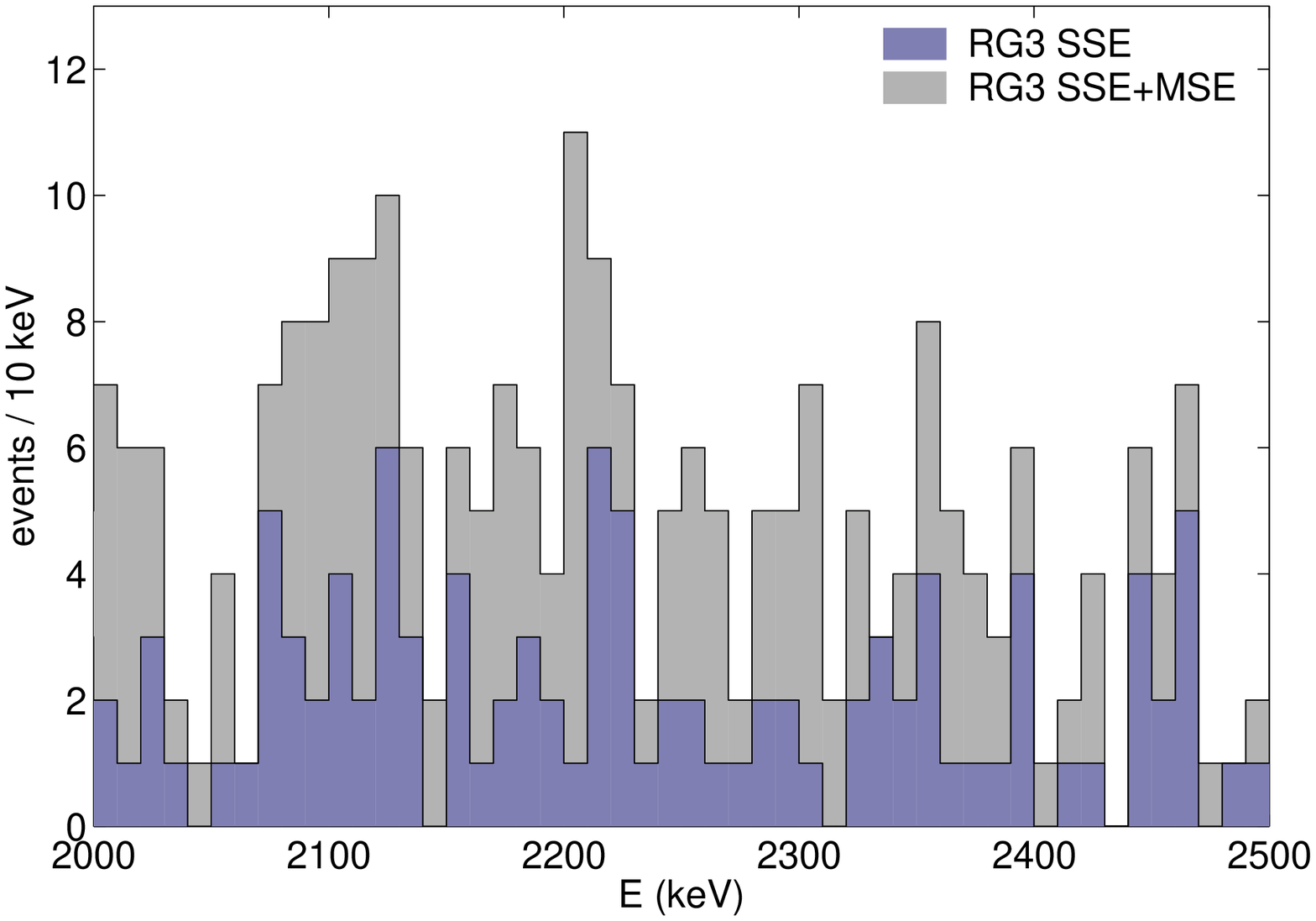}}}
\begin{center}
{\caption {Background spectra before and after the PSD based on
the counting of the number of lobes for detectors RG2 (top) and
RG3 (bottom).} \label{psdlobes}}
\end{center}
\end{figure}

The background reduction technique described above has been
applied to the IGEX data whose pulse shape was recorded (those
events whose pulse shape was not available are conservatively
considered SSE).
%Figure \ref{psdrf} shows the spectra before and after application
%of the PSD technique of the rise-fall time, to the data of
%detectors RG2 and RG3.
Table \ref{psd} summarizes the results (exposure, background
levels in the region of interest from 2 to 2.5 MeV before and
after the PSD, rejection factor) for each detector. Figure
\ref{psdlobes} shows the spectra before and after application of
the PSD technique to the data of detectors RG2 and RG3.
%In much the same way, figure
%\ref{psdlobes} presents the results after applying the technique
%of PSD of counting the number of lobes.
This method results in an efficient rejection %(66.89\% vs 44.98\%)
leading to a background level (in the best case) of 0.10 c/(keV kg
y) in detector RG2, as can be seen in Table \ref{psd}. This value
can be considered as the background limit achieved with the
present PSD technique. The overall final background level of the
set of detectors RG2 and RG3 together turns out to be of 0.10
c/(keV kg y), in the region of interest.

\begin{table}[htb]
\centering \caption {Results of applying the PSD (exposure,
background levels b in the 2-2.5 MeV region before and after the
discrimination and rejection factors).% using two different methods
%based on the rise and fall times (RF) and the counting of the
%number of lobes (L).
} \label{psd} \vskip 0.5 cm
\begin{tabular}{|c||c|c|c|c|} \hline

%& exposure  & b before & b after  & rejection
% &  b after  & rejection
%\\
%
%& & & (RF) & factor & (L) & factor \\
%
%& kg y & c/(keV kg y) & c/(keV kg y) & (RF) \% & c/(keV kg y) &
%(L) \%
%\\ \hline
%
%RG2 & 2.74 & 0.27 & 0.18 & 32.42 & 0.11 & 60.44 \\ \hline
%
%RG3 & 1.90 & 0.26 & 0.09 & 63.79 & 0.06 & 76.54 \\ \hline
%
%total & 4.64 & 0.26 & 0.14 & 44.98 & 0.09 & 66.89 \\ \hline

& exposure  & b before  &  b after  & rejection
\\

& & &  & factor \\

& kg y & c/(keV kg y) & c/(keV kg y) &  (\%)
\\ \hline

RG2 & 2.75 & 0.27 &  0.10 & 62.19 \\ \hline

RG3 & 1.90 & 0.26 &  0.11 & 57.61 \\ \hline

total & 4.65 & 0.26 & 0.10 & 60.36 \\ \hline

\end{tabular}
\end{table}

    The PSD analysis has been applied to only 52.51
mole y out of the total 116.75 mole y (8.87 kg y) accumulated in
the IGEX experiment. In Table \ref{igexdata} the IGEX data
correponding to 8.87 kg.y in $^{76}$Ge in the region between 2020
and 2060 keV, in 2-keV bins, are given, with and without
application of PSD (see also ref. \cite{morales}). The obtained
half-life lower bounds are $T_{1/2}^{0\nu }\geq 1.13\times 10^{25}
y$ for the complete data set and of $T_{1/2}^{0\nu }\geq
1.57\times 10^{25} y$ for the complete data set with application
of PSD to 52.51 mole y. Accordingly, the upper limits on the
neutrino mass parameter are 0.38--1.55~eV for the first data set
and 0.33--1.31~eV for the second data set \cite{igex}. The
uncertainties originate from the spread in the values of the
calculated nuclear matrix elements.
% Despite this fact, the use of PSD allows to
%improve the derived limits at a 90 \% C.L. for the neutrinoless
%half-life (from 1.13 to 1.57$\times 10^{25}$ y) and also for the
%effective neutrino mass .
%tabla de masas????

\begin{table}[ht]
\centering  \caption{IGEX Data bins for 8.87 kg.y in $^{76}$Ge}
\label{igexdata} %\footnotesize
%\vskip 0.5 cm
\begin{tabular}{ccc}
\hline E(low) keV & SSE data set & Complete data set\\\hline
2020 & 2.9 & 3.9 \\

2022 & 9.1 & 10.1 \\

2024 & 2.4 & 4.4\\

2026 & 2.0 & 6.0 \\

2028 & 5.6 & 7.6 \\

2030 & 6.5 & 7.5 \\

2032 & 3.3 & 5.3\\

2034 & 0.6 & 1.6 \\

2036 & 1.0 & 4.0 \\

2038 & 2.0 & 3.0 \\

2040 & 0.5 & 2.5 \\

2042 & 3.5 & 5.5 \\

2044 & 4.0 & 7.0 \\

2046 & 2.7 & 2.7 \\

2048 & 5.3 & 7.3 \\

2050 & 3.4 & 5.4 \\

2052 & 4.6 & 7.6 \\

2054 & 5.0 & 7.0 \\

2056 & 0.6 & 1.6 \\

2058 & 0.1 & 0.1 \\
2060 & 3.3 & 6.3 \\ \hline

Expected counts&13.0&20.3\\

Observed counts&4.1&11.1\\

Upper limit A &3.1 &4.3 \\ (90\%CL) &&\\\hline

ln2.Nt/A&$1.57\times10^{25}$y&$1.13\times10^{25}$y\\
 \hline
%%\normalsize
\end{tabular}
\end{table}

\section{Conclusions}

A Pulse Shape Discrimination technique to reject the radioactive
background in the region in which the double beta decay signal is
expected has been developed and applied to the data collected in
the IGEX experiment, searching for the neutrinoless double beta
decay of $^{76}$Ge. A satisfactory understanding of the pulse
shapes has been achieved. The method described in this paper is
based on the counting of the number of lobes of the pulses, using
a proper filter. It has provided a rejection of $\sim$ 60 \% of
the events in the region of interest, accepting the criterion that
those events having more than two lobes cannot be due to a double
beta decay. Accordingly, the improved background levels provided
by the PSD technique have allowed the improvement of the limits
for the half-life of $^{76}$Ge and consequently, the effective
electron neutrino mass bound.

%indicar algo sobre la limitaci\'{o}n intr\'{\i}nseca del m\'{e}todo estimada??

\section{Acknowledgments} The Canfranc Underground La\-bo\-ra\-to\-ry
is operated by the University of Zaragoza under contract No.
FPA2001-2437. This research was partially founded by the Spanish
Ministry of Science and Technology (MCYT), the US National Science
Foundation and the US Department of Energy. The isotopically
enriched $^{76}$Ge was supplied by the Institute for Nuclear
Research (INR), Moscow, and the Institute for Theoretical and
Experimental Physics (ITEP), Moscow.

\newpage

%\begin{figure}[tb]
%\centerline{ \fboxrule=0cm
% \fboxsep=0cm
%  \fbox{
%\epsfxsize=15cm
%  \epsfbox{risefall.eps}}}
%\begin{center}
%{\caption {Fall vs rise time (both expressed in ns) for calculated
%SSE (top) and background events (bottom). Results are shown for
%detectors RG2 (on the left) and RG3 (on the right).}
%\label{risefall}}
%\end{center}
%\end{figure}

%\begin{figure}[tb]
%\centerline{ \fboxrule=0cm
% \fboxsep=0cm
%  \fbox{
%\epsfxsize=15cm
%  \epsfbox{comparison.eps}}}
%\begin{center}
%{\caption {Examples of the comparison between folded calculated
%SSE and experimental pulses. Events on the left are rejected while
%those on the right are accepted.} \label{comparison}}
%\end{center}
%\end{figure}

%\begin{figure}[tb]
%\centerline{ \fboxrule=0cm
% \fboxsep=0cm
%  \fbox{
%\epsfxsize=12cm %estaba 15, luego 13
%  \epsfbox{psdrf.eps}}}
%\begin{center}
%{\caption {Background spectra before and after the PSD based on
%the rise and fall times for detectors RG2 (top) and RG3 (bottom).}
%\label{psdrf}}
%\end{center}
%\end{figure}

\end{document}